\documentclass[twocolumn,showpacs,aps, prl, floatfix, amsmath]{revtex4}
\usepackage{dcolumn}
\usepackage{bm}
\usepackage{graphicx}
\usepackage{bm}

\begin{document}
\title{Novel thermal effects across first order magnetic transition in Ta doped $HfFe_2$ intermetallic}
\author{Pallab Bag, R Rawat\footnote{E-mail : rrawat@csr.res.in} and P Chaddah}
\affiliation{UGC-DAE Consortium for Scientific Research, University Campus, Khandwa Road, Indore-452001, India.}

\author{P D Babu and V Siruguri}
\affiliation{UGC-DAE Consortium for Scientific Research, R-5 Shed, Bhabha Atomic Research Centre, Mumbai, India.}

\begin{abstract}

Novel thermal effects across the first order antiferromagnetic (AFM) - ferromagnetic (FM) transition in an intermetallic alloy are reported. They show instances of warming when heat is extracted from the sample, and cooling when heat is provided to the sample across AFM-FM transition in Ta doped $HfFe_2$, thus providing indisputable evidence of metastable supercooled AFM and superheated FM states, respectively. Such thermal effects in a magnetic solid prepared from commercially available materials has been reported for the first time. The transition proceeds in multiple steps which is interpreted in the framework of quenched disorder broadening of AFM-FM transition and classical nucleation theory. Measurements in the presence of magnetic field conform to above frame work.
  
\end{abstract}


\maketitle\section{}
First order transitions are characterized by latent heat and transformation from one phase to another phase occurs via nucleation and growth process. If the heat ($\Delta$Q) extracted from the system is less than its total latent heat (L), then only a fraction ($\propto \Delta Q/ L$) of the total system can transform to low temperature phase. However, this fraction cannot be reduced to an arbitrarily small value as the nuclei below a critical size will be unstable. In the classical nucleation theory, the critical size of the stable nuclei (r*) is proportional to ($\gamma_{sl}$ $\times$ $T_C$)/($\Delta T^*$ $\times$ $\Delta L$), where $\gamma_{sl}$ is the interfacial energy, $T_C$ is the equilibrium transition temperature, $\Delta T^*$ is undercooling temperature and $\Delta L$ is the latent heat of the transition of nuclei. Therefore, the transformation from one phase to other phase occurs in a quantum of steps, the size of which is dictated by r*. With increased undercooling, i.e., with increased metastability of supercooled state, r* decreases. Higher the metastability smaller is the perturbation required to transform this state into a stable state. For such transformation during cooling (warming), if the heat removed from (supplied to) the adiabatic system is less than the latent heat of transformation, then system shows warming, when heat is removed from the system (cooling when heat is added to the system). Such thermal effects known as recalescence are commonly observed during liquid to solid transformation (e.g. water to ice transition, solid-liquid transition in NiAl by Kulkarni et al. \cite{Kulkarni1998}, Al-Nb alloys by Munitz et al. \cite{Munitz2000}). However, observation of similar thermal effects across first order magnetic transitions in solids are rare. Generally, first order transitions in solids are broadened due to presence of quenched disorder. For such materials, there exists a spatial distribution of transition temperatures on the length scale of correlation length and for a macroscopic system, it leads to quasi-continuous distribution of transition temperatures leading to apparent gradual change in physical properties \cite{Imry1979, Chaddah201405}. Though locally (on the length scale of correlation length), transition remains discontinuous \cite{Rawat2013a, Roy2004, Soibel2000}. Till date, there are only two examples of magnetic solids with first order phase transition where, such thermal effects has been reported; ultra high purity Er and Dy metals studied by Gschneidner et al.\cite{Pecharsky1996, Gschneidner1997}. Direct observation of such thermal effects was possible due to ultra high purity of the studied system in addition to the lower heat capacity of sample as well as the addenda. Here, we show a similar thermal effect around first order antiferromagnetic (AFM)-ferromagnetic(FM) transition in transition metal alloy (Ta doped $HfFe_2$) prepared from commercial grade (purity $\approx$ 99.99\%) constituent elements. Quenched disorder broadening results in a multiple step transition which is further verified by measurements in the presence of magnetic field.

	The hexagonal parent $HfFe_2$ compound is ferromagnetic at room temperature, which becomes antiferromagnetic with $>$14\% Ta substitution for Hf \cite{Nishihara1982, Nishihara1983}. For  {$\approx$14 to 22\%} Ta substitution, a first order AFM to FM transition with $\approx$1\% volume expansion has been reported \cite{Nishihara1982, Nishihara1983, Rawat2014a}.	The composition studied in the present work, namely $Hf_{0.82}Ta_{0.18}Fe_2$, shows a first order AFM-FM transition around 220 K. The sample is prepared using commercially available Ta and Fe of purity 99.99\% and Hf of purity 99.9\% (exclusive of 2\% Zr). Constituent elements were weighed in their atomic ratio and arc melted three to four times under inert Argon gas atmosphere. Rietveld analysis of powder XRD pattern of as-prepared sample is found to be consistent with hexagonal lattice with a space group P6$_3$/mmc and it showed that sample is single phase with lattice parameters a = 4.9292 \text{\AA} and c = 8.0636 \text{\AA} \cite{Rawat2013b}. Magnetization measurements were carried out using SQUID-VSM and MPMS-VSM, both from M/s. Quantum Design, USA. Heat capacity was measured using a home-made heat capacity set-up based on semi-adiabatic heat pulse method \cite{Rawat2001b, Bag2015} along with 8-Tesla superconducting magnet system from Oxford Instruments, U.K.
	
	The results of magnetization (M) and heat capacity (C$_P$) measured across first order AFM-FM transition temperature are shown in figure 1(a) and (b), respectively. Magnetization data, which are collected during cooling and subsequent warming in the presence of 0.1 Tesla magnetic field, show an AFM-FM transition around 215.5 K and 219.5 K, respectively. In the AFM state, isothermal application of magnetic field can induce an AFM to FM transition; a typical curve is shown in the inset of figure 1(a) at 250~K. It shows an increase of about 1.4 $\mu_B/f.u.$  around 4 Tesla with a narrow hysteresis for increasing and decreasing field cycles. Heat capacity, which is measured during warming, exhibits a sharp peak around 218 K, as highlighted in the main panel of figure 1(b). The right inset shows measured heat capacity (in blue circles) along with sum (red triangles) of linear (with a coefficient 24 mJ/mol~K$^2$)  and Debye contribution (with Debye temperature 355 K). The estimate of latent heat of transition is done from the enthalpy curve calculated from the measured heat capacity as shown in the inset of figure 1(b). The enthalpy curve below and above the transition region is fitted with a linear equation and extrapolated to the transition temperature. The difference between these two curves at transition temperature gives the latent heat of the transition, which is found to be 245 J/mol. These values (entropy change, latent heat, Debye temperature etc.) are in close agreement with those reported by Wada et al.\cite{Wada1993}.

\begin{figure}[hbt]
	\begin{center}
	\includegraphics[width=8 cm]{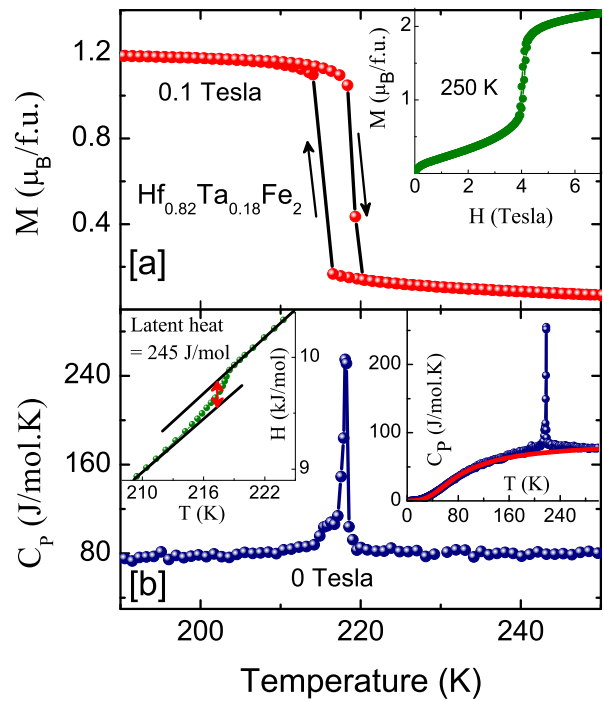}
	\end{center}
	\caption{(color online)\textbf{[a]} Magnetization measured in the presence of 0.1 Tesla during cooling and subsequent warming in $Hf_{0.82}Ta_{0.18}Fe_2$. Inset shows the field induced AFM-FM transition with isothermal application of magnetic field at 250 K. \textbf{[b]} Heat capacity measured during warming in the absence of applied magnetic field across first order AFM-FM transition. Left inset shows the enthalpy variation across first order transition and the right inset shows the measured heat capacity along with sum of the linear term with a coefficient of 24mJ/mol K$^2$ and Debye term with Debye temperature 355 K.}
	\label{Figure1}
\end{figure}

\begin{figure*}[hbt]
	\begin{center}
	\includegraphics[width=16 cm]{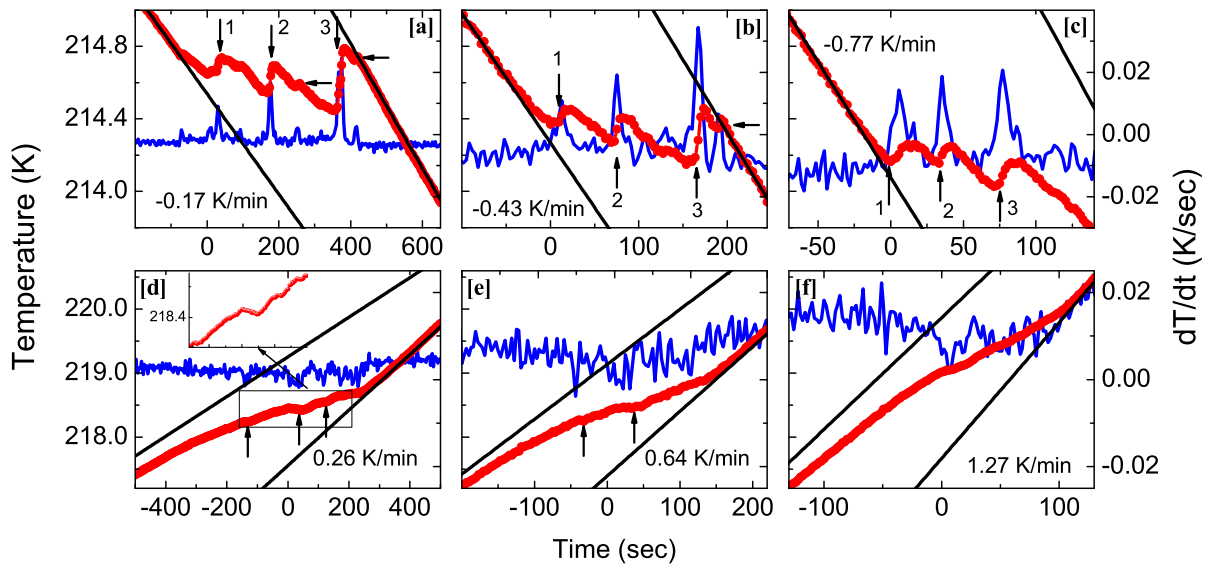}
	\end{center}
	\caption{(color online)Time dependence of sample temperature (right axis) and its derivative (left axis) across first order AFM-FM transition in $Hf_{0.82}Ta_{0.18}Fe_2$ during cooling (top panel) and warming (bottom panel) for various temperature ramp rate in the absence of applied magnetic field. Vertical arrows highlight instances of increase in sample temperature when heat is extracted from the sample and decrease in sample temperature while heat is supplied to the sample. Horizontal arrows mark the instances when jump occurs at higher temperature than the start temperature of preceding jump.}
	\label{Figure2}
\end{figure*}
	 
 Figure 2 shows some typical sample temperature versus time  curves during cooling and warming across AFM-FM transition temperature. These curves were collected using the same set up as used for measuring heat capacity \cite{Rawat2001b, Bag2015}. For these measurements sample holder heater is switched off and heat flow to or from the sample is controlled by varying the radiation shield temperature (or surrounding temperature) at a constant rate. As shown in the top panel of figure 2, during cooling, multiple instances of temperature rise are observed when heat is extracted from the sample around the transition region. The rise in sample temperature when heat is extracted from the sample provides an unambiguous evidence of the transformation of metastable supercooled AFM state into stable FM state. The temperature jumps are found to be as large as 340 mK (figure 2(a), jump no. 3). With increasing ramp rate, these thermal effects shift to lower temperature and step size reduces. Rapid cooling allows deeper supercooling (and hence transformation at lower temperature) which, in the case of metallic glasses, is used to avoid crystallization. The step is expected to vanish if the heat extracted from the system becomes equal or larger than the released latent heat.  Similar temperature versus time scans, taken during warming, are shown in bottom panel of figure 2. In this case, for the lowest heating rate (figure 2(d)), sample temperature decreases (though with much smaller magnitude as highlighted in the inset of figure 2(d)) within the temperature region of first order transition. This observation of cooling when heat is supplied to the system is an evidence of the transformation from superheated FM state to stable AFM state. However, the net changes in sample temperature were found to be much smaller (at the most 50~mK) when compared to that observed during cooling (about 340 mK). For higher ramping rate, these features diminish. This is more akin to water to ice transition, where superheating is found to be absent as melting starts from the surface. Similar to the ice to water transition, in the present system too, the FM to AFM transition is also accompanied with decrease in unit cell volume. The asymmetry in transformation during cooling and heating has been a feature of many magnetoelastic AFM-FM transitions, for example, in case of FeRh, AFM-FM transformation is shown to be asymmetric under certain conditions and explained due to the difference in nucleation and growth mechanisms for AFM to FM and back transformation \cite{Maat2005, Bald2015}. For the present sample also, the M-T measurement (see figure 1(a)) indicates broader AFM-FM transition during warming when compared to cooling.
	
	 The multiple jumps can be interpreted in terms of a broad first order transition due to quenched disorder, where each instance of jump indicates transformation of a different region of the sample. The latent heat associated with these jumps is listed in table 1 and is found to be as high as 49 J/mol for the 340 mK jump (jump no. 3 of figure 2(a)). A comparison with total latent heat of the transition suggests that feature 3 in the cooling curve corresponds to about 20\% of the sample and sum of all jumps accounts for about 44\% of the sample. 	
	
	It is worth mentioning here that multiple jumps in temperature versus time curve in the case of ultra high purity Er have been taken as an indication of the transformation through unknown intermediate states \cite{Gschneidner1997}. However, direct evidence of such intermediate states is yet to be found. In the light of present investigation, the multiple step transition for Er sample can be explained considering disorder broadening. Their heat capacity data of Dy and Er systems also support this interpretation since in the case of Er, a small but non zero width of the transition is evident in the temperature dependence of heat capacity. Whereas for Dy, which showed singular behavior in heat capacity (i.e., no broadening), a single step transition is observed. 
	
	The contention that the observed multiple steps are a consequence of quenched disorder broadening is further tested by studying these thermal effects in the presence of magnetic field. With the application of magnetic field, transition temperature $T_C$ increases and, therefore, critical size of the nuclei is expected to increase. In addition, transition becomes broader and, therefore, temperature difference between consecutive instances of inverse temperature rise is expected to increase. Some typical cooling temperature versus time and heat capacity curves in the presence of 1-4 Tesla magnetic field are shown in figure 3(a-e). As the applied magnetic field increases, the transition temperature is shifted to higher temperature and the spacing between two consecutive jumps increases. The number and magnitude of temperature jumps decrease with magnetic field and at 4 Tesla, no such thermal effects are observed. As the $T_C$ increases, the critical size of the nuclei ($r^*$) increases and this decreased surface to volume ratio results in smaller heat release during recalescence.  These thermal effects are further suppressed due to decreased latent heat of transition with increasing magnetic field as shown in the inset of figure 3(e).  
	
	Another novel feature in the temperature versus time curves shown in figure 2(a) is that the start temperature of the some of the subsequent jumps is not necessarily lower than the start temperature of a preceding jump. Such instances are marked by horizontal arrows in the figure 2(a) and (b). For a disordered broadened first order transition due to quenched disorder, the region with higher supercooling limit will be transformed first during cooling. In this picture, a jump at higher temperature than the preceding one seems to indicate that regions with lower supercooling limit are transforming first. It is possible that the FM region nucleated in the preceding step acts as a nucleation center for the subsequent step resulting in a FM transformation at higher temperature. In the case of La$_{0.5}$Ca$_{0.5}$MnO$_3$, a dramatic increase in resistivity with thermal cycling below transition temperature $T_C$ has been explained due to presence of AFM phase obtained during previous cooling, which results in enhancement of low temperature phase during subsequent warming till T$_C$ \cite{Chaddah2008}.

\begin{table}
\caption{\label{arttype} Temperature rise ($\Delta$T in K) and associated latent heat ($\Delta$L in J/mol) for jumps labeled in figure 2 and 3.}
\begin{ruledtabular}

\begin{tabular}{ccccccc}
Magnetic field (T) / & & Jump No. & \\  
Ramp rate (K/min)	& 1                        & 2                       &3\\
                  & $\Delta T$ / $\Delta L$    & $\Delta T$ / $\Delta L$  & $\Delta T$ / $\Delta L$ \\
0  / 0.2           &0.15 / 26                   & 0.28 / 33                 & 0.34 / 48                \\
0  / 0.4           &0.08 / 32                   & 0.14 / 33                 & 0.31 / 49                \\
0  / 0.8           &0.09 / 32										 & 0.09 / 33                 & 0.13 / 49               \\

1  / 0.2           &0.24 / 33										 & 0.17 / 23                 & 0.13 / 20               \\
2  / 0.2           &0.03 / 32										 & 0.06 / 16                 & 0.06 / 12               \\
3  / 0.2           &0.02 / 18										 & 0.03 / 16                 & - / -               \\
\end{tabular}
\end{ruledtabular}
\end{table}

	\begin{figure}[hbt]
	\begin{center}
	\includegraphics[width=7 cm]{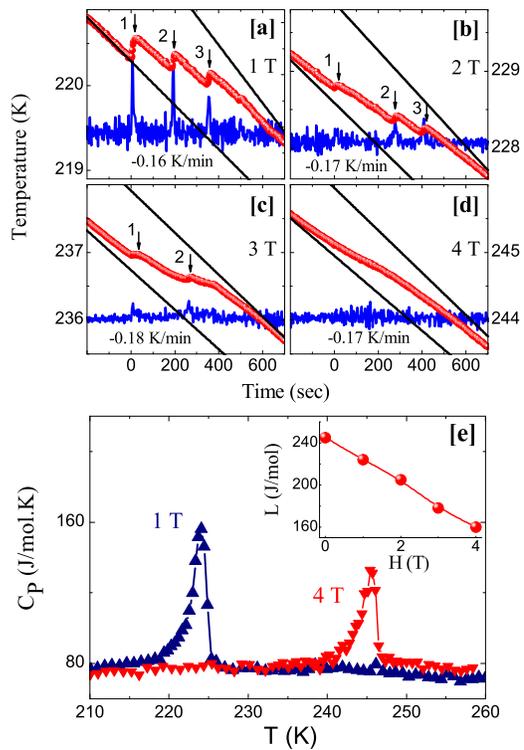}
	\end{center}
	\caption{(color online)\textbf{[a-d]} Time dependence of sample temperature across first order AFM to FM transition in $Hf_{0.82}Ta_{0.18}Fe_2$ during cooling for 1, 2, 3 and 4 Tesla. The instances of temperature rise (indicated by vertical arrows) while heat is extract from the sample vanish with increasing magnetic field. The blue line curves show the derivative of respective curves on a scale of -.015 to +0.045 K/sec. \textbf{[e]} Heat capacity measured during warming in the presence of 1 and 4 Tesla magnetic field . Inset shows the variation of latent heat with increasing magnetic field. }
	\label{Figure3}
\end{figure}

	To conclude, novel thermal effects associated with the transformation of metastable supercooled/superheated to stable states have been observed. This transformation results in warming (cooling) when heat is extracted from (supplied to) the system. This is the first report of such observation in a bulk magnetic solid prepared with commercial grade purity materials, thereby showing a broad first order transition. The transformation takes place in multiple steps, which is interpreted as distribution of transformation temperatures over sample volume as a result of quenched disorder. These features are qualitatively explained in the framework of classical nucleation theory which is tested further with measurement in the presence of magnetic field. Incidentally, there is a striking similarity between AFM-FM transition in the present system and water to ice transition e.g. supercooling but no super heating and higher volume of the low temperature state as compared to the high temperature state etc. If this analogy holds true, then the nucleation of AFM phase is expected to be on surface or grain boundary.

\maketitle\section{Acknowledgement}
R J Choudhary, UGC-DAE Consortium for Scientific Research Indore is acknowledged for magnetization measurements.

{}


\begin{thebibliography}{}

\bibitem[1]{Kulkarni1998} N. S. Kulkarni and K. T. Hong, Metallurgical and Materials Transactions A \textbf{29A}, 2221 (1998).
\bibitem[2]{Munitz2000} A. Munitz, A. B. Gokhlae, and R. Abbaschian, J. Materials Science \textbf{35}, 2263 (2000).
\bibitem[3]{Imry1979} Y. Imry and M. Wortis, Phys. Rev. \textbf{19}, 3580 (1979).
\bibitem[4]{Chaddah201405} P. Chaddah (2014), arXiv-1405.1162.
\bibitem[5]{Rawat2013a} R. Rawat, P. Kushwaha, D. K. Mishra, and V. Sathe, Phys. Rev. B \textbf{87}, 064412 (2013).
\bibitem[6]{Roy2004} S. B. Roy, G. K. Perkins, M. K. Chattopadhyay, A. K. Nigam, K. J. S. Sokhey, P. Chaddah, A. D. Caplin, and L. F. Cohen, Phys. Rev. Lett. \textbf{92}, 147203 (2004).
\bibitem[7]{Soibel2000} A. Soibel, E. Zeldov, M. Rappaport, Y. Myasoedov,T. Tamegai, S. Ooi, M. Konczykowski, and V. B.
Geshkenbeink, Nature \textbf{406}, 282 (2000).
\bibitem[8]{Pecharsky1996} V. K. Pecharsky, J. K.A. Gschneidner, and D. Fort, Scripta Materialia, \textbf{35}, 843 (1996).
\bibitem[9]{Gschneidner1997} K. A. Gschneidner, V. K. Pecharsky, and D. Fort, Phys. Rev. Lett. \textbf{78}, 4281 (1997).
\bibitem[10]{Nishihara1982} Y. Nishihara and Y. Yamaguchi, J. Phys. Soc. Japan \textbf{51}, 1333 (1982).
\bibitem[11]{Nishihara1983} Y. Nishihara and Y. Yamaguchi, J. Phys. Soc. Japan \textbf{52}, 3630 (1983).
\bibitem[12]{Rawat2014a} P. Bag, S. Singh, P. D. Babu, V. Siruguri, and R. Rawat, Physica B \textbf{448}, 50 (2014).
\bibitem[13]{Rawat2013b} R. Rawat, P. Chaddah, P. Bag, P. D. Babu, and V. Siruguri, J. Phys.: Condens Mater \textbf{25}, 066011 (2013).
\bibitem[14]{Rawat2001b} R. Rawat and I. Das, Phys. Rev. B \textbf{64}, 052407 (2001).
\bibitem[15]{Bag2015} P. Bag, V. Singh, and R. Rawat, Rev. Sci. Instrum. \textbf{86}, 056102 (2015).
\bibitem[16]{Wada1993} H. Wada, N. Shimamura, and M. Shiga, Phys. Rev. B \textbf{48}, 102 (1993).
\bibitem[17]{Maat2005} S. Maat, J.-U. Thiele, and E. E. Fullerton, Phys. Rev. B \textbf{72}, 214432 (2005).
\bibitem[18]{Bald2015} C. Baldasseroni, C. Bordel, C. Antonakos, A. Scholl, K. H. Stone, J. B. Kortright, and F. Hellman, J. Phys.: Condens. Matter \textbf{27}, 256001 (2015).
\bibitem[19]{Chaddah2008} P. Chaddah, K. Kumar, and A. Banerjee, Phys. Rev. B \textbf{77}, 100402(R) (2008). 
 
\end{thebibliography}
\end{document}